\documentclass[12pt]{article}
\pdfoutput=1

\usepackage[utf8]{inputenc}
\usepackage[T1]{fontenc}
\usepackage{graphicx}
\usepackage{textcomp}
\usepackage{bm}
\usepackage{cite}
\usepackage{authblk}

\begin{document}
%\linenumbers

\title{Exfoliation of single layer BiTeI flakes}

\author[1,2]{Bálint Fülöp\footnote{fulop.balint@mail.bme.hu, Tel.:+36 1 463 1650}}
\author[3]{Zoltán Tajkov}
\author[4]{János Pető}
\author[4]{Péter Kun}
\author[3]{János Koltai}
\author[5]{László Oroszlány}
\author[1,6]{Endre Tóvári}
\author[7]{Hiroshi Murakawa}
\author[8]{Yoshinori Tokura}
\author[1]{Sándor Bordács}
\author[4]{Levente Tapasztó}
\author[1,6]{Szabolcs Csonka}

\affil[1]{Department of Physics, Budapest University of Technology and Economics, Budafoki út 8, 1111 Budapest, Hungary.}
\affil[2]{MTA-BME Condensed Matter Research Group, Budafoki ut 8, 1111 Budapest, Hungary.}
\affil[3]{Department of Biological Physics, Eötvös Loránd University, Pázmány Péter sétány 1/A, 1117 Budapest , Hungary.}
\affil[4]{Centre for Energy Research, Institute of Technical Physics and Materials Science, 2D Nanoelectronics Lendület Research Group, Konkoly-Thege út 29-33, 1121 Budapest, Hungary}
\affil[5]{Department of Physics of Complex Systems, Eötvös Loránd University, Pázmány Péter sétány 1/A, 1117 Budapest , Hungary.}
\affil[6]{ MTA-BME Lendület Nanoelectronics Research Group, Budafoki út 8, 1111 Budapest, Hungary.}
\affil[7]{Department of Physics, Osaka University, Toyonaka 560-0043, Japan.}
\affil[8]{Department of Applied Physics, The University of Tokyo, Tokyo 113-8656, Japan; RIKEN Center for Emergent Matter Science (CEMS), Wako 351-0198, Japan.}

\date{\today}  
\renewcommand\Affilfont{\itshape\small}
\maketitle
\clearpage

\begin{abstract}
Spin orbit interaction can be strongly boosted when a heavy element is embedded into  an inversion asymmetric crystal field. A simple structure to realize this concept in a 2D crystal contains three atomic layers, a middle one built up from heavy elements  generating strong atomic spin-orbit interaction and two neighboring atomic layers with different electron negativity. BiTeI is a promising candidate for such a 2D crystal, since it contains heavy  Bi layer between Te and I layers. Recently the bulk form of BiTeI attracted considerable attention due to its giant Rashba interaction, however, 2D form of this crystal was not yet created. In this work we report the first exfoliation of single layer BiTeI using a recently developed exfoliation technique on stripped gold. Our combined scanning probe studies and first principles calculations show that SL BiTeI flakes with sizes of 100~\textmu{}m were achieved which are stable at ambient conditions. The giant Rashba splitting and spin-momentum locking of this new member of 2D crystals open the way towards novel spintronic applications and synthetic topological heterostructures. 
\end{abstract}


\section{Introduction}

Recently, the scientific interest and research activity of stacked two dimensional (2D) van der Waals heterostructures have opened a new horizon to engineer materials at the nanoscale. These structures consist of single or few atomic layer thick crystals stacked on top of each other. The first member of the family of 2D materials was graphene, a zero-gap semiconductor, but it includes metals, semiconductors, insulators, furthermore semimetals, superconductors or strongly correlated materials \cite{Geim2013, Koski2013}. For the application of these heterostructures in the field of spintronics and synthetic topological insulators, spin-momentum locking and band inversion are required, respectively, which can be provided by single or few-layer 2D crystals with high spin-orbit interaction (SOI) \cite{Kou2014, Eremeev2015}. Crystals or nanostructures that lack inversion symmetry are good candidates to demonstrate strong SOI \cite{LaShell1996, ast2007, Koroteev2004, Nitta1997}. Among these, the polar semiconductor BiTeI is a very promising candidate due to its giant Rashba splitting \cite{Bahramy2011}, but so far no fabrication of single layer BiTeI (SL BiTeI, one triplet of Te-Bi-I atomic layers) flakes has been reported.

\begin{figure}[t]
\centering
\includegraphics[height=0.4\textwidth]{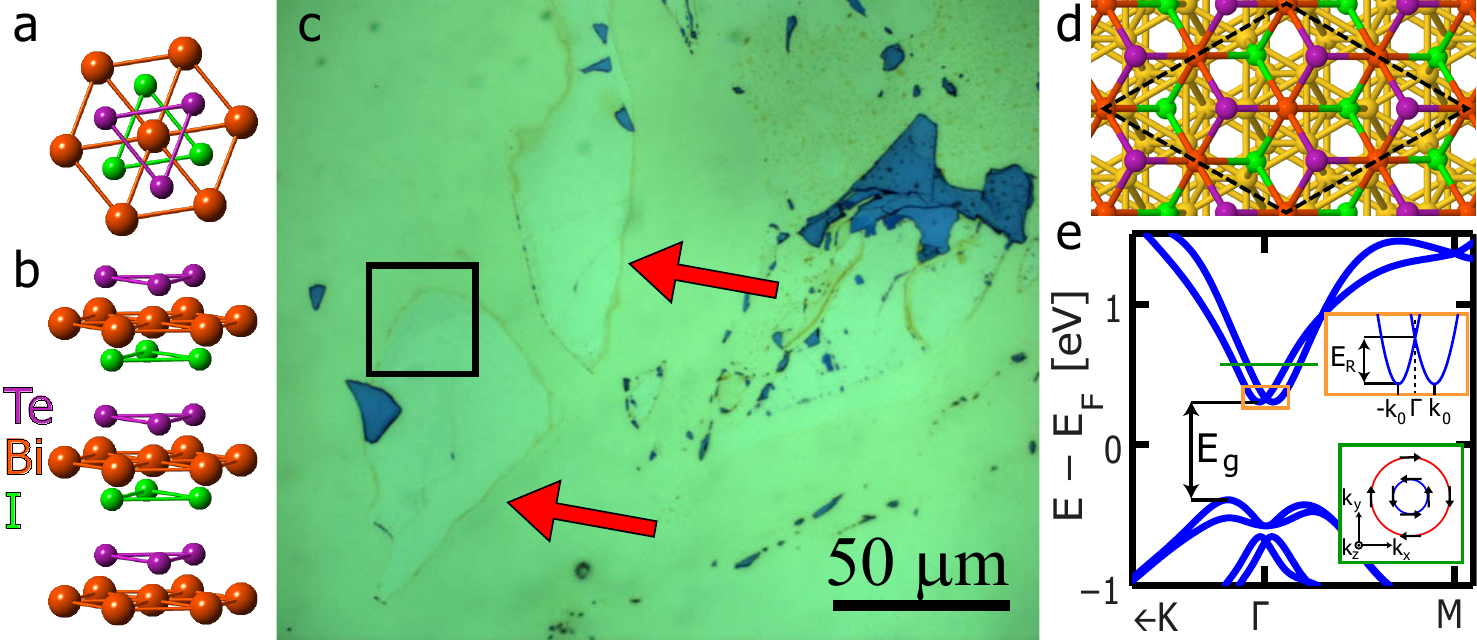}
\caption{(a-b) Top and side view of the structure of the BiTeI crystals, respectively. The asymmetric stacking of the Te, Bi, I layers breaks the inversion symmetry and leads to the polar structure. (c) Channel-selective contrast enhanced optical micrograph of the BiTeI flake on stripped gold surface after sonication. The green background correspond to the Au substrate, the blue flakes are the thick BiTeI crystals. The light green patches with brown borders marked by red arrows are the SL BiTeI flakes. Brown glue residues are also visible spluttered on the surface in the top right corner of the image. The black square marks the location of the AFM measurements discussed in Fig.~\ref{fig:AFM}. (d) Illustration of the theoretically investigated BiTeI\,+\,Au system (I faced). The unit cell (depicted with black dashed lines) consists of 66 atom in total (54 gold atoms, 4 atoms of Bi, Te and I each). (e) Dispersion relation (energy versus momentum relationship) in freestanding SL BiTeI along the K~--~$\Gamma$~--~M points. The band gap $E_\textrm{g}$, the momentum offset $k_0$ and the Rashba Energy $E_\textrm{R}$ are also indicated. Lower inset: schematic representation of the spin-momentum locking effect on the vertical cut of the dispersion relation along the marked green section.}
\label{fig:micrOpt}
\end{figure}

BiTeI is a member of a new class of polar crystals with layered structure, the class of ternary bismuth tellurohalides BiTeX (X = I, Br, Cl), that recently attracted considerable attention \cite{Kulbachinskii2012, Martin2016, Sakano2013, Landolt2012, Ishizaka2011, Kou2014, Eremeev2015, Nechaev2017, Eremeev2017, Bahramy2012, Ohmura2017, Eremeev2012, Shevelkov1995, Kanou2013, Sakano2012, Monserrat2017, Crepaldi2012, Butler2014, Lee2011, Demko2012, Bordacs2013, Ogawa2013, Fiedler2014, Kohsaka2015, PRB.88.081104}. The key component is Bi, which as a heavy element has a strong atomic SOI. Its triangular lattice layer is asymmetrically stacked between a Te and an I (or Br or Cl) layer (see Fig.~\ref{fig:micrOpt}a-b) \cite{Ishizaka2011}. The Bi layer along with tellurium forms a positively charged (BiTe)$^+$ layer with similar geometry to metallic bismuth, whereas the I$^-$ layer is negatively charged \cite{Shevelkov1995}. This polar structure, the narrow band gap and the same orbital character of the bands at the top valence band and the bottom conduction band lead to the appearence of a giant Rashba spin splitting \cite{Kulbachinskii2012, Martin2016}. 

The Rashba effect leads to a Hamiltonian $H_\mathrm{R} = \alpha \cdot \bm{\sigma} (\mathbf{n} \times \mathbf{k})$, where $\alpha$ is a coupling constant,  $\bm\sigma$ is a vector of the Pauli matrices acting on the elecron spin, \textbf{n} is a unit vector pointing out of the 2D plane of the crystal and \textbf{k} is the in-plane electron wave vector \cite{Bychkov1984}.  The corresponding band structure is shown at Fig.~\ref{fig:micrOpt}e. The Rashba effect dictates a spin momentum locking as the lower inset shows and it induces an energy shift  between opposite spin direction, which is described by the Rashba-energy, $E_\mathrm{R}$. $E_\mathrm{R}$ is exceptionally high for BiTeI, for bulk BiTeI $E_\mathrm{R} \approx 110$~meV \cite{Sakano2013, Bahramy2011}. This is four times higher than energy scale of the room temperature thermal fluctuations ($k_\mathrm{B} T = 25$~meV), two orders of magnitude higher than the spin splitting measured on a conventional InGaAs/InAlAs semiconductor interface, or measured on the surface of Ag(111) or Au(111) \cite{Ishizaka2011, ast2007}. Only extremely sensitive surface structures like Bi atoms on Ag surface existing only in ultra high vacuum have higher Rashba energy \cite{ast2007}. This built-in giant spin-orbit interaction makes BiTeI an attractive, novel component in van der Waals heterostructures. 

%Crystal structure and micrOpt picture

Recently several theoretical works proposed combination of BiTeI with other 2D crystals. According to first principles calculations of BiTeI/gra\-phene \cite{Kou2014, Tajkov2017} (and BiTeCl/graphene \cite{Eremeev2014}) heterostructures, the strong Rashba interaction of BiTeI is expected to exert a significant influence on the Dirac electrons of graphene resulting in a nontrivial band structure, which paves the way for a new class of robust artificial topological insulators with vast possible application in spintronics. While in graphene the 2D Dirac states originally do not have spin-momentum locking, it is also possible to combine SL BiTeI with topological insulators that host 2D helical Dirac states, thus resulting a prototype to examine various more complex spin transport effect depending on the coupling strength between these systems \cite{Eremeev2015}. Furthermore, a pair of inversely stacked SL BiTeI itself also expected to show topological insulating behaviour\cite{Nechaev2017,Eremeev2017}. This shows the versatility of possible applications of SL BiTeI and the increasing demand for its production. However, no experimental demonstration of SL BiTeI has been reported yet. Previous studies only focussed on bulk properties \cite{Shevelkov1995, Ishizaka2011, Martin2016, Kanou2013, Sakano2012, Landolt2012, Monserrat2017, Bahramy2012, Crepaldi2012, Butler2014, Lee2011, Demko2012, Bordacs2013, Ogawa2013, PRB.88.081104}.

The standard mechanical exfoliation technique, using adhesive tape pres\-sed and released on a SiO$_2$ substrate, is successful for many 2D materials such as graphene and hexagonal boron nitride (hBN) because of the strong adhesion between these crystals and the SiO$_2$ surface\cite{Koenig2011}. However, according to our early experience, in the case of BiTeI mechanical exfoliation results in only 50-100~nm thin flakes with few micrometers of lateral dimensions and very low yield. In the literature, BiTeI thin films of thickness from 70~nm to 10~\textmu{}m have been fabricated from polycrystalline powder using flash evaporation and their mechanical \cite{Onopko1972a} and electrical \cite{Onopko1972b} properties have been investigated, but these polycrystalline samples are far thicker than a single layer of the crystal. 

\section{Results and Discussion}

In the current work SL BiTeI flakes were exfoliated with a novel method using freshly cleaved Au(111) substrate\cite{Magda2015} that yields SL BiTeI flakes of lateral dimensions up to 100~\textmu{}m. The underlying principle of the method that Te and I bonds to Au stronger than the cohesive energy between two BiTeI layers. Thus the last layer of BiTeI remains on the gold surface when a bulk BiTeI crystal comes into contact with Au surface and then detaches due to sonication. To understand the exfoliation process we also performed density functional theory (DFT) based calculations to obtain the geometry and binding energies. In the following we present our optical microscopy, STM and AFM measurements made on three samples obtained by the fabrication method presented in the Methods section, then compare them to the results of our thoretical calculations. 

A typical optical micrograph of a sample surface after exfoliation is shown in Fig.\,\ref{fig:micrOpt}c. In this picture, due to the channel-selective contrast enhancement, the Au substrate is green and thick BiTeI crystals that stayed on the surface after the sonications are blue. Some of the thick BiTeI crystals that detached during sonication leave behind light green patches of several tens of microns lateral size, like the pair marked by red arrows in Fig.~\ref{fig:micrOpt}c. In the following we focus on these regions and show that SL BiTeI covers almost continously these patches.

%atomic structure STM figure
\begin{figure}[t]
\centering
\includegraphics[width=\textwidth]{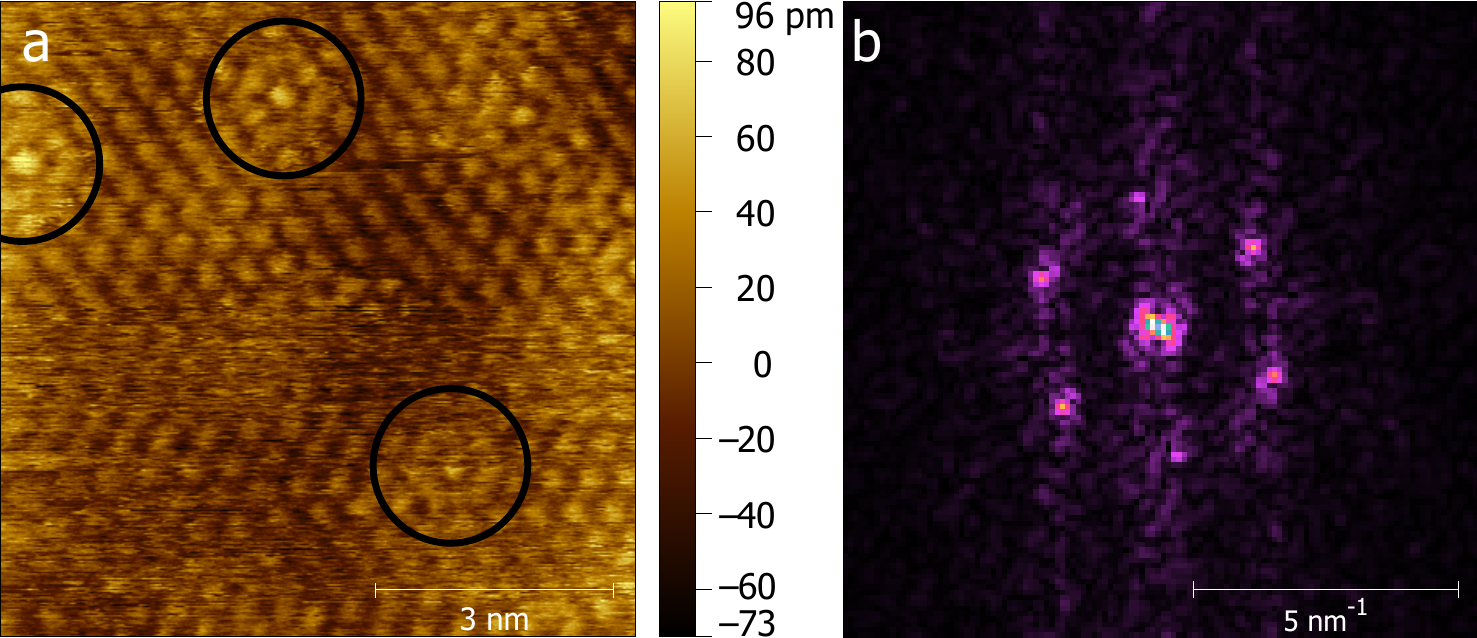}
\caption{(a) STM image of the atomic structure of the SL BiTeI flake in a 8~nm~$\times$~8~nm window. The observed trigonal pattern corresponds to the crystal structure of the BiTeI. Black circles mark defect sites. (b) 2D FFT spectrum of the structure. The peak positions correspond to 4.1~\AA $\pm$ 0.2~\AA, which is in agreement with the measured bulk lattice constants of BiTeI in Refs. \cite{Kulbachinskii2012, Sankar2014}.}
\label{fig:atomic}
\end{figure}

As a next step, atomic resolution STM measurements were made on these patches. Room temperature, ambient STM measurements revealed a trigonal atomic pattern at the surface (see Fig.~\ref{fig:atomic}a), similar to the bulk crystal structure of BiTeI. To precisely measure the periodicity of the observed trigonal pattern, we performed a two dimensional Fourier transform (see Fig.~\ref{fig:atomic}b) and measured the positions of the maxima yielding a periodicity $4.2\pm 0.2$~\AA{}. This is in agreement with the bulk lattice parameter of BiTeI in the layer plane according to previous reports (4.3~\AA \cite{Kulbachinskii2012,Sankar2014}), and is significantly different from the gold lattice parameter in the (111) direction (2.9~\AA\cite{Nie2012}). Thus, one can conclude that after sonication BiTeI is still present on the substrate.   Analyzing further Fig.~\ref{fig:atomic}a, one can also find defects with bright apperance (see black circles). Comparing with bulk defect states of BiTeI \cite{Fiedler2014, Kohsaka2015}, the absence of pronounced threefold symmetry and the small height of $<20$~pm suggest that they are antisites.

%cloudy STM picture
\begin{figure}[t]
\centering

\includegraphics[width=\textwidth]{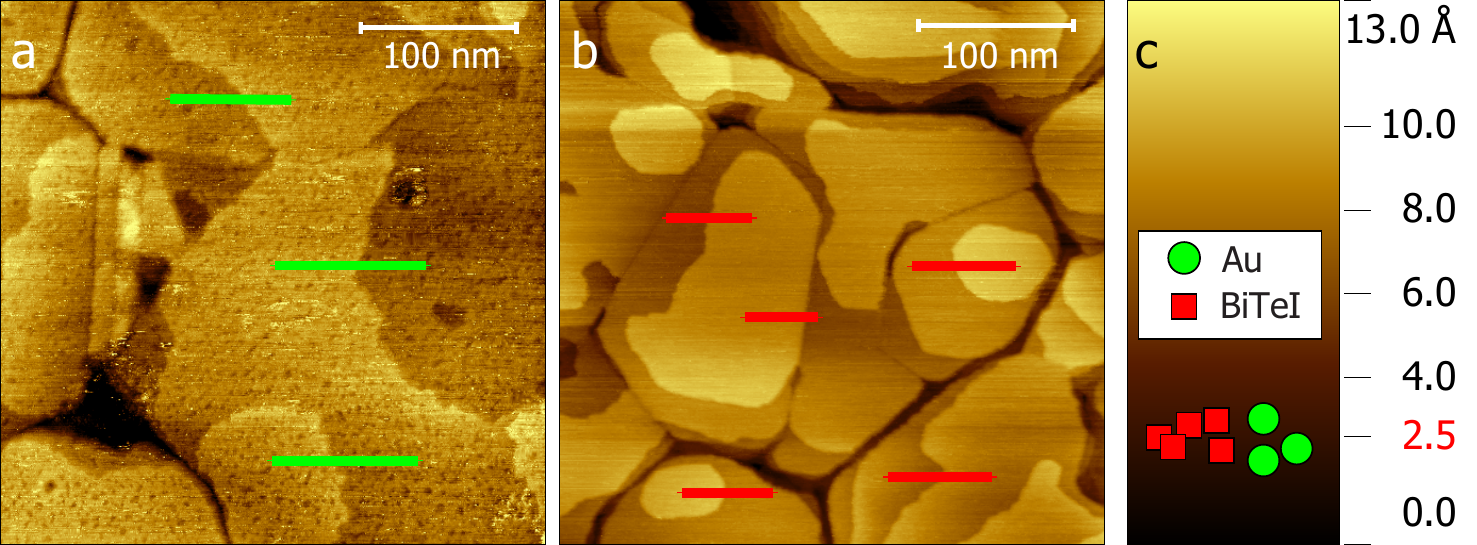}
\caption{ Comparison of surface topologies. (a), (b) STM image of a typical pure Au (111) and BiTeI covered surface, respectively. The presence of BiTeI alters the surface significantly. The steps are more pronounced while the terraces are smaller in lateral extent leading to more steps present on the same scan size. Scan size is 350~nm~$\times$~350~nm. The green and red segments mark cuts along which step height measurements were performed and indicated on the heat map. (c) Heat map used on the maps (a-b) and marks of heights of the steps measured along the marked segments on panel (a) (green) and panel (b) (red). The step heights  measured on the BiTeI cloudy pattern do not differ from those measured on pure gold. The STM images were taken at ambient conditions.}
\label{fig:STM}
\end{figure}

\begin{figure}[t]
\centering
\includegraphics[width=\textwidth]{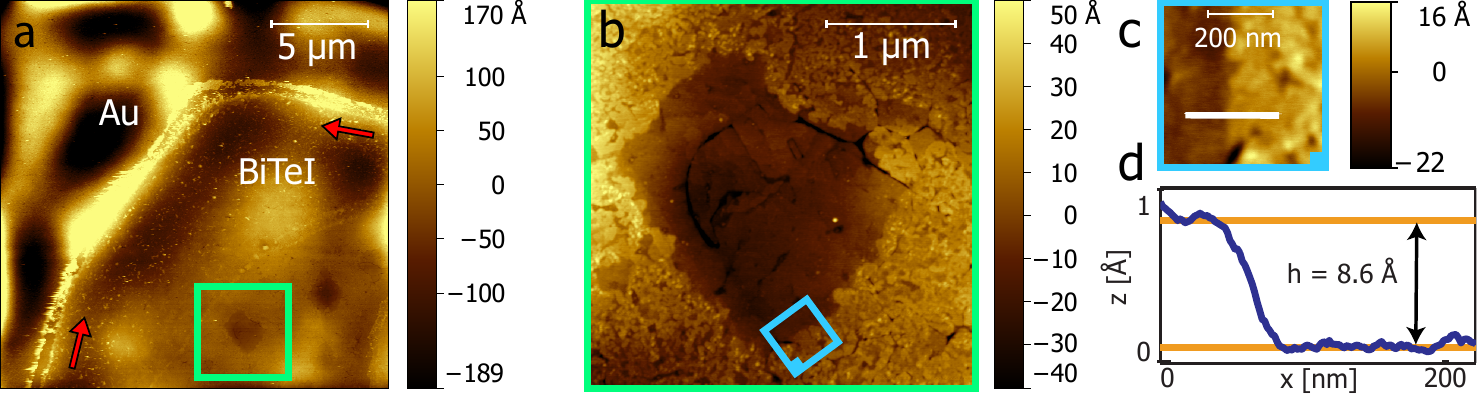}
\caption{(a)AFM image of the location marked in Fig.1(a) by the black square. The bright curved line marked by the red arrows is the top-left border of the flake. The green square marks one of the hole locations where the BiTeI layer is missing and the pure stripped gold surface is reproduced, the region presented at panel (b). Its sharp contour and pronounced color contrast is different from the dark and light spots across the image caused by the height of the contamination accumulated on the border during the applied polynomial background correction. (b) In the outer region the cloudy pattern is visible which is characteristic of the BiTeI surfaces whereas the inner region shows landscape typical for stripped gold surface. The blue square marks the location of the layer thickness measurement shown in (c-d). (c) Example of the layer thickness measurements and (d) the line plot of the corresponding cut.}
\label{fig:AFM}
\end{figure}

When zooming out for larger area scans, we found that the surface topology of the BiTeI flakes is also different from the pure Au substrate at the lengthscale of 1~\textmu{}m (see Fig.~\ref{fig:STM}). The Au surfaces are well known to consist of large (111) terraces separated by single or multiple steps along (112) or (110) directions measuring 2.5~\AA\ step heights (see Fig.\,\ref{fig:STM}a) \cite{PhysRevLett.60.120, Barth1990}. The areas which were covered by BiTeI crystal before sonication show characteristically different landscape (see Fig.\,\ref{fig:STM}b). Let us call it ``cloudy'' texture in the following. In this pattern one can find distinctive terraces of atomically flat regions similar to that of Au (111) faces, but in this case the boundaries of the terraces are more ragged, their size is smaller, contains multiple terraces on top of each other which are not present in the case of Au. The two surface structures are also well distinguishable in larger scan ranges as a later AFM image shows (see Fig.~\ref{fig:AFM}b). The trigonal atomic structure presented in Fig.~\ref{fig:atomic}a can be generally found anywhere where the cloudy pattern is visible and it is never present on the pure gold surface. 

In order to identify whether the terraces are related to atomic steps of BiTeI or to the substrate, we measured the height of dozens of the terraces using linecuts in the scan direction, some of which are marked on Fig.~\ref{fig:STM}a-b using green and red segments for the clean Au and the BiTeI covered surfaces, respectively. Step heights corresponding to the marked segments are indicated by markers on the height colorbar in Fig.~\ref{fig:STM}c. We found that the measured step height of the cloudy pattern corresponds to that of the Au (111) steps of 2.5~\AA\cite{PhysRevLett.60.120, Barth1990}, and is largely different from the bulk lattice parameters of BiTeI in the out-of-plane direction (6.5~\AA \cite{Kulbachinskii2012}). This suggests that these steps do not mark the border of BiTeI monolayers but the gold terraces of the surface that the BiTeI layer follows closely. To identify the thickness of the covering BiTeI layer we have to search for an area where the flake ends on the optical image or the cloudy pattern ends in STM (called ``border regions'').

In Fig.~\ref{fig:AFM}a we show an AFM overview image of the investigated flake depicted in Fig.~\ref{fig:micrOpt}a. The bright curvy line connecting the two red arrows is the border of the flake, i.e. the top-left corner is the Au surface whereas the larger part of the image bounded by the bright line is covered by BiTeI. At the border of the flake a thick and broad pile of accumulated material can be found. The width of this line is rather large, 2-4~\textmu{}m, which is in the same length scale as the surface roughness of the Au substrate, thus the measurement of the layer thickness across the border can not be performed reliably. Therefore we looked for holes in the flake such as the one marked by the green square. In these holes the surface is deeper and the texture is different. 

After zooming on the area marked by the green square (Fig. \ref{fig:AFM}b) one can recognize the cloudy pattern of BiTeI (same as in STM measurement at Fig.~\ref{fig:STM}b) in the outer region whereas in the hole the original Au surface is present, which corroborates the visual impression that the BiTeI layer is missing in the hole. On the border of these holes the accumulated contamination is not present therefore it is possible to zoom in further (see Fig.~\ref{fig:AFM}c) and measure directly the step height at the border (Fig.~\ref{fig:AFM}d). We investigated the step height at various positions around the border of the hole similar to Fig.~\ref{fig:AFM}c, measuring a couple of linecuts using various PID control parameters. The measured step heights are in the range of $8.5 \pm 1.2$~\AA, which is close to the bulk lattice parameters of BiTeI in the out-of-plane direction (6.5~\AA \cite{Kulbachinskii2012} or 6.8~\AA \cite{Sankar2014}). Thus we conclude that the measured step height corresponds to a single layer step, and the regions showing the cloudy pattern are covered by a monolayer of BiTeI crystal. The small mismatch of the measured height and lattice parameter can be attributed to the fact that height measurement on different substrate could deviate slightly in AFM profiles\cite{nemes2008}. Thus, our findings indicate that SL BiTeI can indeed be separated by stripped gold exfoliation technique, thus, it is a powerful method to produce large size SL 2D crystals from materials beyond graphene and TMDCs \cite{Magda2015}. The BiTeI covered surface was analyzed several weeks after exfoliation as well and the atomic structure of the BiTeI layer (as in Fig.~\ref{fig:atomic}) was still presented which shows the long term environmental stability of the BiTeI monolayer. This finding is also remarkable, since TMDCs containing Te are usually unstable in ambient conditions on the Au surface in the time scale of several hours according to our previous experiments.

%Hole in the flake AFM picture

To support the experimental results, we performed first principles calculations. First, we investigated the electronic structure of the freestanding single layer and compared to the calculated properties of the bulk reproduced from Ref.~\cite{Bahramy2011}. Our calculated band structure of the freestanding SL BiTeI is presented in Fig.~\ref{fig:micrOpt}e, which provided a band gap of $E_\mathrm{g} = 740$~meV, and a Rashba energy of $E_\mathrm{R} = 35$~meV.

As a next step SL BiTeI on Au surface was investigated (see Fig.~\ref{fig:micrOpt}d). Geometry optimization left the Au surface largely unaltered, while it introduced a small buckling of 0.1~\AA\  in the BiTeI layer irrespective of Te or I was facing the Au substrate. The binding energies of the relevant bonds are listed in Table~\ref{tab:binding_energies}. The energy relations clearly indicate that both the Te-faced and the I-faced BiTeI binds stronger to the Au surface than to another BiTeI layer (see the first 3 rows), which is in agreement with the experimental findings that a SL BiTeI remains at the Au surface after sonication. As a reference, we also included the bonding energies between the constituents of a single BiTeI layer which are much higher than the previous ones. Thus, it is highly unlikely that the BiTeI layer can be cleaved between Bi-Te or Bi-I planes, leaving only a part of the single layer on the substrate. This result further supports that the multiple terraces at the cloudy pattern (see Fig.~\ref{fig:STM}b) are not related to BiTeI but the underlying gold surface. On the other hand, the  strong adhesion between the BiTeI and the Au substrate can be also a reason why the surface structure of the Au is significantly different under BiTeI coverage (cloudy pattern): the strong bonds can block the Au atoms which have otherwise high surface mobility; and also force the BiTeI to follow the terraces. This surface reconstruction is likely to be induced by the relative high temperature ($T \approx90$~C$^\circ$) that the sample was exposed to during fabrication.

\begin{table}[ht]
\centering
\begin{tabular}{rl}
	Bond & Binding energy \\
	\hline
  	Au$^6$--I--Bi--Te $\rightarrow$ Au$^6$ + BiTeI & 681~meV \\
   	Au$^6$--Te--Bi--I $\rightarrow$ Au$^6$ + BiTeI & 969~meV \\
   	BiTeI--BiTeI $\rightarrow$ BiTeI + BiTeI & 543~meV \\
	Te--Bi--I $\rightarrow$ Te--Bi + I & 2.74~eV \\
   	I--Bi--Te $\rightarrow$ I--Bi + Te & 3.64~eV\\
\end{tabular}
\caption{Binding energies calculated by PBE + Grimme method. The values represent the energy needed to break the bond as marked by the arrow. The results indicate that the BiTeI tends to stick to the Au surface stronger than to another layer BiTeI, both Te or I-faced. The internal bonds are much stronger than the interlayer binding energies, making it unlikely to leave only part of the BiTeI layer on the substrate.}
\label{tab:binding_energies}
\end{table}

\begin{figure}[t]
\includegraphics[width=\textwidth]{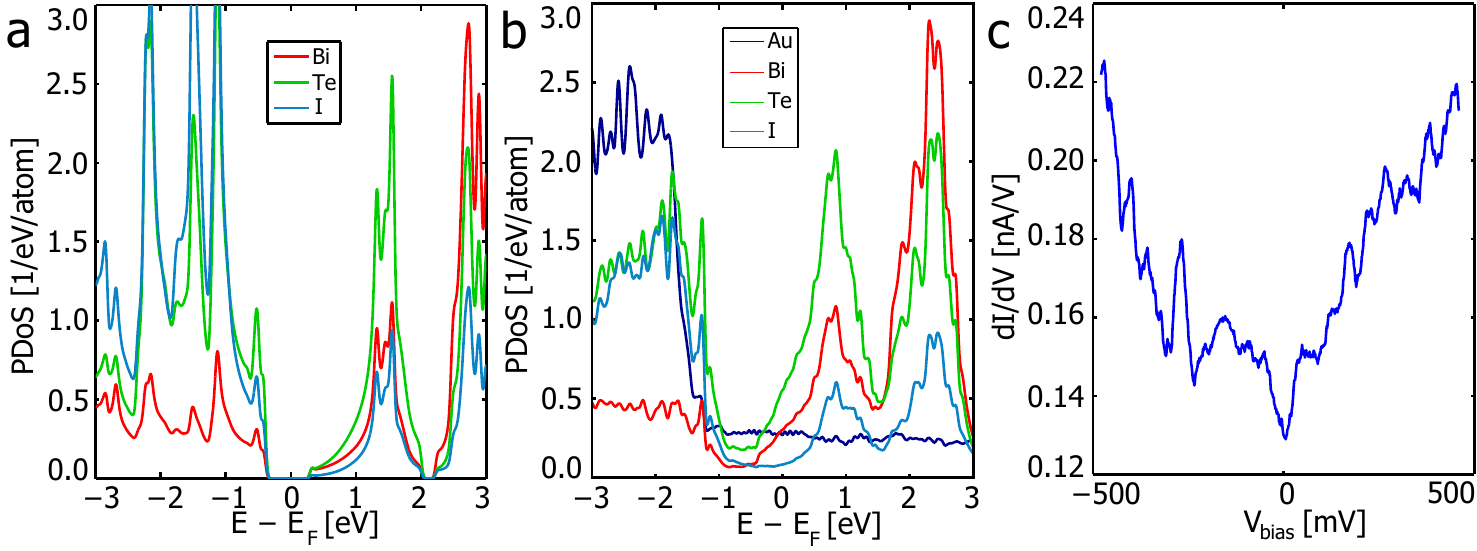}
\caption{(a) Calculated partial density of states (PDOS) of the freestanding SL BiTeI. (b) PDOS of the SL BiTeI on Au, Te-side (i.e. Au$^6$-Te-Bi-I). (c) Numerical derivative of a measured I-V curve on a SL BiTeI atomic structure. The derivative does never go to zero which is in agreement with the calculated PDOS curves of (b).}
\label{fig:PDOS}
\end{figure}

To further characterize the obtained flakes, we calculated the partial DOS (PDOS) on the constituents with and without the presence of the Au substrate (see Fig.~\ref{fig:PDOS}a-b, respectively). In the case of the freestanding SL BiTeI one can see a gap of 0.76~eV where the PDOS is zero for all the components, therefore we expect the SL BiTeI to show insulating behavior in this regime (see Fig.~\ref{fig:PDOS}a and Fig.~\ref{fig:micrOpt}e). In the whole investigated range of $\pm3$~eV, the position of the most prominent peaks in the PDoS are very similar for Bi, Te and I as well. One can observe that in the negative energy range the Te and I have more contribution while the part of te Bi orbitals increases towards higher energies. In the middle range, between 1 and 2~eV, the Te has the highest PDOS. 

Meanwhile, when the SL BiTeI is placed on the Au substrate, the PDOS change significantly (see Fig.~\ref{fig:PDOS}b). The influence of the Au substrate can depend on the orientation of the crystal structure of BiTeI with respect to the Au substrate, whether it is placed on the substrate in the order Au$^6$-Te-Bi-I (Te-side) or Au$^6$-I-Bi-Te (I-side). After evaluation of the PDOS structures of the two cases, we indeed see a difference between the two sides, but it is minor compared to the freestanding case. Therefore, we only illustrate the effect of the Au substrate using the plot of the PDOS of Te-side (Fig.~\ref{fig:PDOS}b). One can clearly see the additional states of the Au substrate in the negative range and the smeared peaks of every component in the positive range. Most importantly, the PDOS never reaches zero for any of the components, meaning that the gap disappears due to the strong hybridization with the Au substrate. Therefore insulating behavior is not expected in the case of the SL BiTeI placed on the stripped gold substrate.

To confirm this prediction, tunneling current -- voltage characteristics measurements were performed. A typical example of a dI/dV curve measured on the cloudy pattern is shown on Fig.~\ref{fig:PDOS}c. The conductance never reaches zero and there is no clear signature of a gap in the measurement, which is in agreement with the calculated PDOS. However, even without hybridization the presence of the Au substrate can add a strong, featureless background to the measured dI/dV characteristics by letting electrons tunnel through the BiTeI layer directly into the Au substrate.

%PDOS figure

\section{Conclusion}

We demonstrated for the first time that single layer flakes can be realized from the giant Rashba spin-orbit material BiTeI. Stripped gold exfoliation technique provides an efficient way to produce flakes with a size of 100~\textmu{}m, which are stable at ambient condition for at least several weeks. We showed that after the exfoliation the position of the BiTeI flakes can be identified by simple optical microscope. Atomic resolution STM measurements confirmed the presence of BiTeI layer on the gold surface by resolving the in-plane crystal structure of BiTeI. AFM measurements showed that the flakes cover large areas continuously with only few holes. Step height measurements across the edges of these holes confirmed that the flake thickness corresponds to SL BiTeI. Our first principles calculations also supported the formation of SL BiTeI due to the strong bonding between Te and I to Au substrate. Moreover, BiTeI strongly hybridize with the Au substrate, which results in finite DoS in the gap in accordance with the differential conduction measurements. The first exfoliation of SL BiTeI adds a new member to the possible building blocks of van der Waals heterstructures with giant Rashba spin splitting, and thereby opens the way to engineer novel 2D heterostructures with special spin based functionality or topological protection. 

\section*{Methods}

\subsection*{Fabrication}
\label{sec:fabr}

Gold layers of 100~nm thickness were grown epitaxially on mica prior to exfoliation. Before use the mica-Au interface was freshly cleaved, thus obtaining large area Au (111) surfaces where the exfoliation took place. Bulk BiTeI crsytals were grown by the Bridgman method as described in Ref.\cite{Ishizaka2011}. Thick BiTeI flakes were prepared on scotch tape by consecutive folding several times and transferred on the Au (111) surface using a thermal release tape. After the removal of the thermal release tape by heating the sample on a hot plate up to 90~$^\circ$C, the sample was mildly sonicated in room temperature acetone. The sonication causes some of the thick BiTeI flakes to fall off the substrate leaving only SL BiTeI pieces on the surface. The SL BiTeI flakes can be found by investigating these areas using channel-selective contrast-enhanced optical microscopy as depicted in Fig.\,\ref{fig:micrOpt}c.

\subsection*{Optical and scanning probe microscopy}

A Zeiss Axio Imager optical microscope was used for the optical investigation of the samples. We took 2.5x magnification pictures before and after the sonication and compared them to find areas where the bulk crystals fell off during the process (not shown in this paper). We found that the position of the single layer BiTeI flakes can be localized using channel-selective contrast-enhanced optical microscope images  with 100x microscope lens (see Fig.~\ref{fig:micrOpt}c).

Scanning Tunneling Microscopy (STM) measurements were performed on a Nanoscope E instrument using standard Pt-Ir 90\%-10\% tips created by mechanical shearing. High resolution 2D maps were scanned at a setpoint of 3~nA at 5~mV, large area 2D maps were scanned at 1~nA and 200~mV. The differential conduction measurements were obtained by measuring I-V characteristics at 1~nA and 200~mV, then a numerical differentiation was applied to the average of dozens of individual measurements. Atomic Force Microscope (AFM) was used in tapping mode. The used Bruker Multimode 8 Nanocope V. AFM instrument has larger scanning range than the STM device but sufficient resolution in the z axis to resolve atomic layer thicknesses. All STM and AFM measurements were performed under ambient conditions.

\subsection*{First principles calculations}

The density functional theory calculations were performed for two configurations: the first model consists of a freestanding SL BiTeI cell, the second is the SL BiTeI placed on the Au substrate. In the case of the freestanding single layer, the slabs were separated by a considerable thick vacuum of at least $18.5$~\AA. For the SL BiTeI on Au, we considered the following geometry (depicted in Fig. \ref{fig:micrOpt}d) for calculating the binding energies and the density of states (DoS) of SL BiTeI on the Au surface: the $2\times 2$ supercell of the SL BiTeI was placed on 6 layers of $3\times 3$ supercell of (111) Au both Te and I faced (latter is shown in Fig. \ref{fig:micrOpt}d). The mismatch of the two lattices is $1.3\%$ only, which is albeit expected to alter the calculated properties slightly but it is not expected to affect our conclusions. The slabs were also separated by a vacuum of at least $18.5$~\AA.

The geometry optimization, the binding energies and the DOS were calculated both using the projector augmented-wave method as 
implemented in the VASP Package \cite{kresse_1993,kresse_1996} and in the linear combination of atomical orbitals method as implemented in the SIESTA package \cite{siesta_2002}.

During optimization and the calculation of the binding energies the SOI was neglected as previous works showed that these quantities are barely sensitive to it \cite{Kou2014,Tajkov2017}. The parameters used in the VASP calculation were the following: the plane-wave cutoff was 500~eV, the Brilloiun zone was sampled with the $12\times 12 \times 1$ Gamma-centered Monckhorst-Pack set. In the geometry optimization we kept the lattice constant of the gold fixed and relaxed all the atoms until the atomic forces fell below 3~meV/\AA.

Next, the binding energies were calculated by substracting the total energies of the fully relaxed, separated fractions from the total energy of the fully relaxed compound. To test the validity of the applied method the binding energies of graphite layers (57~meV/atom) were considered and found to be in good agreement with experimental data in the literature ($62\pm 5\,\rm meV/atom$) \cite{zacharia_2004}. In order to take the van der Waals interactions into account we deployed the DFT-D3 Grimme corrections \cite{grimme_2010} for the Perdew–Burke–Ernzerhof (PBE) functionals \cite{pbe_1996}.

The parameters of the SIESTA computation were the following: the mesh grid cutoff was $300\,\rm Ry$, the Brilloiun zone was sampled with the $5\times 5\times 2$ Gamma-centered Monckhorst-Pack set. The force tolerance in coordinate optimization was $20$~meV/\AA. During relaxation we deployed the PBE functional. We used the pseudopotentials optimized by Rivero et al. \cite{rivero_2015}. The basis size was the double-zeta polarized. 

After relaxation a self-consistent single-point calculation was done with SOI included \cite{VASP_SOC,siesta_SOC_2004}. The sisl tool \cite{zerothi_sisl} was used to extract the partial density of states from SIESTA calculations, sampling the Brillouin zone with a $70\times 70\times 1$ $k$-points set. 

\section*{Acknowledgements}
This work was financially supported by the Flag-ERA iSpinText project (NN118996), the "Lendület" program of the Hungarian Academy of Sciences, the Hungarian National Research, Development and Innovation Office (NKFIH) grants no. K115608, K108676, and K115575;  the Hungarian Research Funds OTKA PD 111756 and  FK 124723. We acknowledges the financial support of the National Research, Development and Innovation Office of Hungary via the National Quantum Technologies Program NKP-2017-00001. S.B., L.O. and J.K. acknowledges the Bolyai program of the Hungarian Academy of Sciences. J.P., P.K. and L.T. acknowledge financial support from ERC StG Nanofab2D and OTKA grant K108753.

\section*{Contributions}
B. F.,  E. T., S. B. and S. C. conceived the project. H. M. and Y. T.  grew the BiTeI crystals, B. F. and J. P. worked out the exfoliation procedure, did the STM studies and analyzed the data with L. T., E. T. and S. C.  P. K.  supported the AFM analysis. Z. T.,  J. K. and L. O. did the calculations. B. F. and S. C. wrote the paper with all authors   contributing to the discussion and preparation of the manuscript.

\providecommand{\WileyBibTextsc}{}
\let\textsc\WileyBibTextsc
\providecommand{\othercit}{}
\providecommand{\jr}[1]{#1}
\providecommand{\etal}{~et~al.}


\begin{thebibliography}{[10]}

\bibitem{Geim2013}% article
 \textsc{A.\,K. Geim} and  \textsc{I.\,V. Grigorieva}\iffalse Van der waals
  heterostructures\fi,
 \jr{Nature} \textbf{499}(7459), 419--425 (2013).


\bibitem{Koski2013}% article
 \textsc{K.\,J. Koski} and  \textsc{Y.~Cui}\iffalse The new skinny in
  two-dimensional nanomaterials\fi,
 \jr{ACS Nano} \textbf{7}(5), 3739--3743 (2013).


\bibitem{Kou2014}% article
 \textsc{L.~Kou},  \textsc{S.\,C. Wu},  \textsc{C.~Felser},
  \textsc{T.~Frauenheim},  \textsc{C.~Chen},  and  \textsc{B.~Yan}\iffalse
  Robust 2d topological insulators in van der waals heterostructures\fi,
 \jr{ACS Nano} \textbf{8}(10), 10448--10454 (2014),
PMID: 25226453.


\bibitem{Eremeev2015}% article
 \textsc{S.\,V. Eremeev},  \textsc{S.\,S. Tsirkin},  \textsc{I.\,A. Nechaev},
  \textsc{P.\,M. Echenique},  and  \textsc{E.\,V. Chulkov}\iffalse New
  generation of two-dimensional spintronic systems realized by coupling of
  rashba and dirac fermions\fi,
 \jr{Scientific Reports} \textbf{5}(August), 12819 (2015).


\bibitem{LaShell1996}% article
 \textsc{S.~LaShell},  \textsc{B.\,A. McDougall},  and
  \textsc{E.~Jensen}\iffalse Spin splitting of an au(111) surface state band
  observed with angle resolved photoelectron spectroscopy\fi,
 \jr{Phys. Rev. Lett.} \textbf{77}(Oct), 3419--3422 (1996).


\bibitem{ast2007}% article
 \textsc{C.\,R. Ast},  \textsc{J.~Henk},  \textsc{A.~Ernst},
  \textsc{L.~Moreschini},  \textsc{M.\,C. Falub},  \textsc{D.~Pacil\'e},
  \textsc{P.~Bruno},  \textsc{K.~Kern},  and  \textsc{M.~Grioni}\iffalse Giant
  spin splitting through surface alloying\fi,
 \jr{Phys. Rev. Lett.} \textbf{98}(May), 186807 (2007).


\bibitem{Koroteev2004}% article
 \textsc{Y.\,M. Koroteev},  \textsc{G.~Bihlmayer},  \textsc{J.\,E. Gayone},
  \textsc{E.\,V. Chulkov},  \textsc{S.~Bl\"ugel},  \textsc{P.\,M. Echenique},
  and  \textsc{P.~Hofmann}\iffalse Strong spin-orbit splitting on bi
  surfaces\fi,
 \jr{Phys. Rev. Lett.} \textbf{93}(Jul), 046403 (2004).


\bibitem{Nitta1997}% article
 \textsc{J.~Nitta},  \textsc{T.~Akazaki},  \textsc{H.~Takayanagi},  and
  \textsc{T.~Enoki}\iffalse Gate control of spin-orbit interaction in an
  inverted
  i${\mathrm{n}}_{0.53}$g${\mathrm{a}}_{0.47}$as/i${\mathrm{n}}_{0.52}$a${\mathrm{l}}_{0.48}$as
  heterostructure\fi,
 \jr{Phys. Rev. Lett.} \textbf{78}(Feb), 1335--1338 (1997).


\bibitem{Bahramy2011}% article
 \textsc{M.\,S. Bahramy},  \textsc{R.~Arita},  and  \textsc{N.~Nagaosa}\iffalse
  Origin of giant bulk rashba splitting: Application to bitei\fi,
 \jr{Phys. Rev. B} \textbf{84}(Jul), 041202 (2011).


\bibitem{Kulbachinskii2012}% article
 \textsc{V.\,A. Kulbachinskii},  \textsc{V.\,G. Kytin},  \textsc{A.\,A.
  Kudryashov},  \textsc{A.\,N. Kuznetsov},  and  \textsc{A.\,V.
  Shevelkov}\iffalse On the electronic structure and thermoelectric properties
  of bitebr and bitei single crystals and of bitei with the addition of bii3
  and cui\fi,
 \jr{Solid State Chemistry and Materials Science of Thermoelectric Materials}
  \textbf{193}(September), 154--160 (2012).


\bibitem{Martin2016}% article
 \textsc{C.~Martin},  \textsc{A.\,V. Suslov},  \textsc{S.~Buvaev},
  \textsc{A.\,F. Hebard},  \textsc{P.~Bugnon},  \textsc{H.~Berger},
  \textsc{A.~Magrez},  and  \textsc{D.\,B. Tanner}\iffalse Experimental
  determination of the bulk rashba parameters in bitebr\fi,
 \jr{EPL (Europhysics Letters)} \textbf{116}(5), 57003 (2016).


\bibitem{Sakano2013}% article
 \textsc{M.~Sakano},  \textsc{M.\,S. Bahramy},  \textsc{A.~Katayama},
  \textsc{T.~Shimojima},  \textsc{H.~Murakawa},  \textsc{Y.~Kaneko},
  \textsc{W.~Malaeb},  \textsc{S.~Shin},  \textsc{K.~Ono},
  \textsc{H.~Kumigashira},  \textsc{R.~Arita},  \textsc{N.~Nagaosa},
  \textsc{H.\,Y. Hwang},  \textsc{Y.~Tokura},  and
  \textsc{K.~Ishizaka}\iffalse Strongly spin-orbit coupled two-dimensional
  electron gas emerging near the surface of polar semiconductors\fi,
 \jr{Phys. Rev. Lett.} \textbf{110}(Mar), 107204 (2013).


\bibitem{Landolt2012}% article
 \textsc{G.~Landolt},  \textsc{S.\,V. Eremeev},  \textsc{Y.\,M. Koroteev},
  \textsc{B.~Slomski},  \textsc{S.~Muff},  \textsc{T.~Neupert},
  \textsc{M.~Kobayashi},  \textsc{V.\,N. Strocov},  \textsc{T.~Schmitt},
  \textsc{Z.\,S. Aliev},  \textsc{M.\,B. Babanly},  \textsc{I.\,R.
  Amiraslanov},  \textsc{E.\,V. Chulkov},  \textsc{J.~Osterwalder},  and
  \textsc{J.\,H. Dil}\iffalse Disentanglement of surface and bulk rashba spin
  splittings in noncentrosymmetric bitei\fi,
 \jr{Phys. Rev. Lett.} \textbf{109}(Sep), 116403 (2012).


\bibitem{Ishizaka2011}% article
 \textsc{K.~Ishizaka},  \textsc{M.\,S. Bahramy},  \textsc{H.~Murakawa},
  \textsc{M.~Sakano},  \textsc{T.~Shimojima},  \textsc{T.~Sonobe},
  \textsc{K.~Koizumi},  \textsc{S.~Shin},  \textsc{H.~Miyahara},
  \textsc{A.~Kimura},  \textsc{K.~Miyamoto},  \textsc{T.~Okuda},
  \textsc{H.~Namatame},  \textsc{M.~Taniguchi},  \textsc{R.~Arita},
  \textsc{N.~Nagaosa},  \textsc{K.~Kobayashi},  \textsc{Y.~Murakami},
  \textsc{R.~Kumai},  \textsc{Y.~Kaneko},  \textsc{Y.~Onose},  and
  \textsc{Y.~Tokura}\iffalse Giant rashba-type spin splitting in bulk bitei\fi,
 \jr{Nat Mater} \textbf{10}(7), 521--526 (2011).


\bibitem{Nechaev2017}% article
 \textsc{I.\,A. Nechaev},  \textsc{S.\,V. Eremeev},  \textsc{E.\,E.
  Krasovskii},  \textsc{P.\,M. Echenique},  and  \textsc{E.\,V.
  Chulkov}\iffalse Quantum spin hall insulators in centrosymmetric thin films
  composed from topologically trivial bitei trilayers\fi,
 \jr{Scientific Reports} \textbf{7}(March), 43666 (2017).


\othercit
\bibitem{Eremeev2017}% misc
 \textsc{S.\,V. Eremeev},  \textsc{I.\,A. Nechaev},  and  \textsc{E.\,V.
  Chulkov},
2d and 3d topological phases in bite$x$ compounds, 2017.


\bibitem{Bahramy2012}% article
 \textsc{M.\,S. Bahramy},  \textsc{B.\,J. Yang},  \textsc{R.~Arita},  and
  \textsc{N.~Nagaosa}\iffalse Emergence of non-centrosymmetric topological
  insulating phase in bitei under pressure\fi,
 \jr{Nature Communications} \textbf{3}(February), 679 (2012).


\bibitem{Ohmura2017}% article
 \textsc{A.~Ohmura},  \textsc{Y.~Higuchi},  \textsc{T.~Ochiai},
  \textsc{M.~Kanou},  \textsc{F.~Ishikawa},  \textsc{S.~Nakano},
  \textsc{A.~Nakayama},  \textsc{Y.~Yamada},  and  \textsc{T.~Sasagawa}\iffalse
  Pressure-induced topological phase transition in the polar semiconductor
  bitebr\fi,
 \jr{Phys. Rev. B} \textbf{95}(Mar), 125203 (2017).


\bibitem{Eremeev2012}% article
 \textsc{S.\,V. Eremeev},  \textsc{I.\,A. Nechaev},  \textsc{Y.\,M. Koroteev},
  \textsc{P.\,M. Echenique},  and  \textsc{E.\,V. Chulkov}\iffalse Ideal
  two-dimensional electron systems with a giant rashba-type spin splitting in
  real materials: Surfaces of bismuth tellurohalides\fi,
 \jr{Phys. Rev. Lett.} \textbf{108}(Jun), 246802 (2012).


\bibitem{Shevelkov1995}% article
 \textsc{A.\,V. Shevelkov},  \textsc{E.\,V. Dikarev},  \textsc{R.\,V.
  Shpanchenko},  and  \textsc{B.\,A. Popovkin}\iffalse Crystal structures of
  bismuth tellurohalides bitex (x = cl, br, i) from x-ray powder diffraction
  data\fi,
 \jr{Journal of Solid State Chemistry} \textbf{114}(2), 379--384 (1995).


\bibitem{Kanou2013}% article
 \textsc{M.~Kanou} and  \textsc{T.~Sasagawa}\iffalse Crystal growth and
  electronic properties of a 3d rashba material, bitei, with adjusted carrier
  concentrations\fi,
 \jr{Journal of Physics: Condensed Matter} \textbf{25}(13), 135801 (2013).


\bibitem{Sakano2012}% article
 \textsc{M.~Sakano},  \textsc{J.~Miyawaki},  \textsc{A.~Chainani},
  \textsc{Y.~Takata},  \textsc{T.~Sonobe},  \textsc{T.~Shimojima},
  \textsc{M.~Oura},  \textsc{S.~Shin},  \textsc{M.\,S. Bahramy},
  \textsc{R.~Arita},  \textsc{N.~Nagaosa},  \textsc{H.~Murakawa},
  \textsc{Y.~Kaneko},  \textsc{Y.~Tokura},  and  \textsc{K.~Ishizaka}\iffalse
  Three-dimensional bulk band dispersion in polar bitei with giant rashba-type
  spin splitting\fi,
 \jr{Phys. Rev. B} \textbf{86}(Aug), 085204 (2012).


\othercit
\bibitem{Monserrat2017}% misc
 \textsc{B.~Monserrat} and  \textsc{D.~Vanderbilt},
Temperature dependence of the bulk rashba splitting in the bismuth
  tellurohalides, 2017.


\bibitem{Crepaldi2012}% article
 \textsc{A.~Crepaldi},  \textsc{L.~Moreschini},  \textsc{G.~Aut\`es},
  \textsc{C.~Tournier-Colletta},  \textsc{S.~Moser},  \textsc{N.~Virk},
  \textsc{H.~Berger},  \textsc{P.~Bugnon},  \textsc{Y.\,J. Chang},
  \textsc{K.~Kern},  \textsc{A.~Bostwick},  \textsc{E.~Rotenberg},
  \textsc{O.\,V. Yazyev},  and  \textsc{M.~Grioni}\iffalse Giant ambipolar
  rashba effect in the semiconductor bitei\fi,
 \jr{Phys. Rev. Lett.} \textbf{109}(Aug), 096803 (2012).


\bibitem{Butler2014}% article
 \textsc{C.\,J. Butler},  \textsc{H.\,H. Yang},  \textsc{J.\,Y. Hong},
  \textsc{S.\,H. Hsu},  \textsc{R.~Sankar},  \textsc{C.\,I. Lu},
  \textsc{H.\,Y. Lu},  \textsc{K.\,H.\,O. Yang},  \textsc{H.\,W. Shiu},
  \textsc{C.\,H. Chen},  \textsc{C.\,C. Kaun},  \textsc{G.\,J. Shu},
  \textsc{F.\,C. Chou},  and  \textsc{M.\,T. Lin}\iffalse Mapping polarization
  induced surface band bending on the rashba semiconductor bitei\fi,
 \jr{NatComm} \textbf{5}(June), 4066 (2014).


\bibitem{Lee2011}% article
 \textsc{J.\,S. Lee},  \textsc{G.\,A.\,H. Schober},  \textsc{M.\,S. Bahramy},
  \textsc{H.~Murakawa},  \textsc{Y.~Onose},  \textsc{R.~Arita},
  \textsc{N.~Nagaosa},  and  \textsc{Y.~Tokura}\iffalse Optical response of
  relativistic electrons in the polar bitei semiconductor\fi,
 \jr{Phys. Rev. Lett.} \textbf{107}(Sep), 117401 (2011).


\bibitem{Demko2012}% article
 \textsc{L.~Demk\'o},  \textsc{G.\,A.\,H. Schober},  \textsc{V.~Kocsis},
  \textsc{M.\,S. Bahramy},  \textsc{H.~Murakawa},  \textsc{J.\,S. Lee},
  \textsc{I.~K\'ezsm\'arki},  \textsc{R.~Arita},  \textsc{N.~Nagaosa},  and
  \textsc{Y.~Tokura}\iffalse Enhanced infrared magneto-optical response of the
  nonmagnetic semiconductor bitei driven by bulk rashba splitting\fi,
 \jr{Phys. Rev. Lett.} \textbf{109}(Oct), 167401 (2012).


\bibitem{Bordacs2013}% article
 \textsc{S.~Bord\'acs},  \textsc{J.\,G. Checkelsky},  \textsc{H.~Murakawa},
  \textsc{H.\,Y. Hwang},  and  \textsc{Y.~Tokura}\iffalse Landau level
  spectroscopy of dirac electrons in a polar semiconductor with giant rashba
  spin splitting\fi,
 \jr{Phys. Rev. Lett.} \textbf{111}(Oct), 166403 (2013).

\bibitem{Ogawa2013}% article
 \textsc{N.~Ogawa},  \textsc{M.\,S. Bahramy},  \textsc{H.~Murakawa},
  \textsc{Y.~Kaneko},  and  \textsc{Y.~Tokura}\iffalse Magnetophotocurrent in
  bitei with rashba spin-split bands\fi,
 \jr{Phys. Rev. B} \textbf{88}(Jul), 035130 (2013).


\bibitem{Fiedler2014}% article
 \textsc{S.~Fiedler},  \textsc{L.~El-Kareh},  \textsc{S.\,V. Eremeev},
  \textsc{O.\,E. Tereshchenko},  \textsc{C.~Seibel},  \textsc{P.~Lutz},
  \textsc{K.\,A. Kokh},  \textsc{E.\,V. Chulkov},  \textsc{T.\,V. Kuznetsova},
  \textsc{V.\,I. Grebennikov},  \textsc{H.~Bentmann},  \textsc{M.~Bode},  and
  \textsc{F.~Reinert}\iffalse Defect and structural imperfection effects on the
  electronic properties of bitei surfaces\fi,
 \jr{New Journal of Physics} \textbf{16}(7), 075013 (2014).


\bibitem{Kohsaka2015}% article
 \textsc{Y.~Kohsaka},  \textsc{M.~Kanou},  \textsc{H.~Takagi},
  \textsc{T.~Hanaguri},  and  \textsc{T.~Sasagawa}\iffalse Imaging ambipolar
  two-dimensional carriers induced by the spontaneous electric polarization of
  a polar semiconductor bitei\fi,
 \jr{Phys. Rev. B} \textbf{91}(Jun), 245312 (2015).


\bibitem{PRB.88.081104}% article
 \textsc{C.\,R. Wang},  \textsc{J.\,C. Tung},  \textsc{R.~Sankar},
  \textsc{C.\,T. Hsieh},  \textsc{Y.\,Y. Chien},  \textsc{G.\,Y. Guo},
  \textsc{F.\,C. Chou},  and  \textsc{W.\,L. Lee}\iffalse Magnetotransport in
  copper-doped noncentrosymmetric bitei\fi,
 \jr{Phys. Rev. B} \textbf{88}(Aug), 081104 (2013).


\bibitem{Bychkov1984}% article
 \textsc{Y.\,A. Bychkov} and  \textsc{E.\,I. Rashba}\iffalse Properties of a 2d
  electron gas with lifted spectral degeneracy\fi,
 \jr{Journal of Experimental and Theoretical Physics} \textbf{39}(2), 78--82
  (1984).


\bibitem{Tajkov2017}% article
 \textsc{Z.~Tajkov},  \textsc{D.~Visontai},  \textsc{P.~Rakyta},
  \textsc{L.~Oroszlány},  and  \textsc{J.~Koltai}\iffalse Transport properties
  of graphene-bitei hybrid structures\fi,
 \jr{Physica Status Solidi C} pp.\,1700215--n/a (2017),
1700215.


\bibitem{Eremeev2014}% article
 \textsc{S.\,V. Eremeev},  \textsc{I.\,A. Nechaev},  \textsc{P.\,M. Echenique},
   and  \textsc{E.\,V. Chulkov}\iffalse Spin-helical dirac states in graphene
  induced by polar-substrate surfaces with giant spin-orbit interaction: a new
  platform for spintronics\fi,
 \jr{Scientific Reports} \textbf{4}(November), 6900 (2014).


\bibitem{Koenig2011}% article
 \textsc{S.\,P. Koenig},  \textsc{N.\,G. Boddeti},  \textsc{M.\,L. Dunn},  and
  \textsc{J.\,S. Bunch}\iffalse Ultrastrong adhesion of graphene membranes\fi,
 \jr{Nat Nano} \textbf{6}(9), 543--546 (2011).


\bibitem{Onopko1972a}% article
 \textsc{L.\,V.\,O. et~al.}\iffalse The method of preparation structure and
  mechanical properties of bitel thin films\fi,
 \jr{Izv VUZ Fiz} \textbf{11}, 117--120 (1972).


\bibitem{Onopko1972b}% article
 \textsc{L.\,V.\,O. et~al.}\iffalse Electro-physical properties of bitel thin
  films\fi,
 \jr{Izv VUZ Fiz} \textbf{11}, 120--122 (1972).


\bibitem{Magda2015}% article
 \textsc{G.\,Z. Magda},  \textsc{J.~Pető},  \textsc{G.~Dobrik},
  \textsc{C.~Hwang},  \textsc{L.\,P. Biró},  and
  \textsc{L.~Tapasztó}\iffalse Exfoliation of large-area transition metal
  chalcogenide single layers\fi,
 \jr{Scientific Reports} \textbf{5}(October), 14714 (2015).

\bibitem{Sankar2014}% article
 \textsc{R.~Sankar},  \textsc{I.~Panneer~Muthuselvam},  \textsc{C.\,J. Butler},
   \textsc{S.\,C. Liou},  \textsc{B.\,H. Chen},  \textsc{M.\,W. Chu},
  \textsc{W.\,L. Lee},  \textsc{M.\,T. Lin},  \textsc{R.~Jayavel},  and
  \textsc{F.\,C. Chou}\iffalse Room temperature agglomeration for the growth of
  bitei single crystals with a giant rashba effect\fi,
 \jr{CrystEngComm} \textbf{16}(37), 8678--8683 (2014).


\bibitem{Nie2012}% article
 \textsc{S.~Nie},  \textsc{N.\,C. Bartelt},  \textsc{J.\,M. Wofford},
  \textsc{O.\,D. Dubon},  \textsc{K.\,F. McCarty},  and
  \textsc{K.~Thürmer}\iffalse Scanning tunneling microscopy study of graphene
  on au(111): Growth mechanisms and substrate interactions\fi,
 \jr{Phys. Rev. B} \textbf{85}(20), 205406 (2012).


\bibitem{PhysRevLett.60.120}% article
 \textsc{R.\,C. Jaklevic} and  \textsc{L.~Elie}\iffalse
  Scanning-tunneling-microscope observation of surface diffusion on an atomic
  scale: Au on au(111)\fi,
 \jr{Phys. Rev. Lett.} \textbf{60}(Jan), 120--123 (1988).


\bibitem{Barth1990}% article
 \textsc{J.\,V. Barth},  \textsc{H.~Brune},  \textsc{G.~Ertl},  and
  \textsc{R.\,J. Behm}\iffalse Scanning tunneling microscopy observations on
  the reconstructed au(111) surface: Atomic structure, long-range
  superstructure, rotational domains, and surface defects\fi,
 \jr{Phys. Rev. B} \textbf{42}(Nov), 9307--9318 (1990).


\bibitem{nemes2008}% article
 \textsc{P.~Nemes-Incze},  \textsc{Z.~Osv{\'a}th},  \textsc{K.~Kamar{\'a}s},
  and  \textsc{L.~Bir{\'o}}\iffalse Anomalies in thickness measurements of
  graphene and few layer graphite crystals by tapping mode atomic force
  microscopy\fi,
 \jr{Carbon} \textbf{46}(11), 1435--1442 (2008).


\bibitem{kresse_1993}% article
 \textsc{G.~Kresse} and  \textsc{J.~Hafner}\iffalse \textit{{Ab} initio}
  molecular dynamics for liquid metals\fi,
 \jr{Physical Review B} \textbf{47}(1), 558 (1993).


\bibitem{kresse_1996}% article
 \textsc{G.~Kresse} and  \textsc{J.~Furthm{\"u}ller}\iffalse Efficient
  iterative schemes for ab initio total-energy calculations using a plane-wave
  basis set\fi,
 \jr{Physical review B} \textbf{54}(16), 11169 (1996).


\bibitem{siesta_2002}% article
 \textsc{J.\,M. Soler},  \textsc{E.~Artacho},  \textsc{J.\,D. Gale},
  \textsc{A.~García},  \textsc{J.~Junquera},  \textsc{P.~Ordejón},  and
  \textsc{D.~Sánchez-Portal}\iffalse The siesta method for ab initio order- n
  materials simulation\fi,
 \jr{Journal of Physics: Condensed Matter} \textbf{14}(11), 2745 (2002).


\bibitem{zacharia_2004}% article
 \textsc{R.~Zacharia},  \textsc{H.~Ulbricht},  and  \textsc{T.~Hertel}\iffalse
  Interlayer cohesive energy of graphite from thermal desorption of
  polyaromatic hydrocarbons\fi,
 \jr{Phys. Rev. B} \textbf{69}(Apr), 155406 (2004).


\bibitem{grimme_2010}% article
 \textsc{S.~Grimme},  \textsc{J.~Antony},  \textsc{S.~Ehrlich},  and
  \textsc{H.~Krieg}\iffalse A consistent and accurate ab initio parametrization
  of density functional dispersion correction ({DFT}-{D}) for the 94 elements
  {H}-{Pu}\fi,
 \jr{The Journal of Chemical Physics} \textbf{132}(15), 154104 (2010).


\bibitem{pbe_1996}% article
 \textsc{J.\,P. Perdew},  \textsc{K.~Burke},  and
  \textsc{M.~Ernzerhof}\iffalse Generalized gradient approximation made
  simple\fi,
 \jr{Phys. Rev. Lett.} \textbf{77}(Oct), 3865--3868 (1996).


\bibitem{rivero_2015}% article
 \textsc{P.~Rivero},  \textsc{V.\,M. García-Suárez},
  \textsc{D.~Pereñiguez},  \textsc{K.~Utt},  \textsc{Y.~Yang},
  \textsc{L.~Bellaiche},  \textsc{K.~Park},  \textsc{J.~Ferrer},  and
  \textsc{S.~Barraza-Lopez}\iffalse Systematic pseudopotentials from reference
  eigenvalue sets for {DFT} calculations\fi,
 \jr{Computational Materials Science} \textbf{98}(February), 372--389 (2015).


\bibitem{VASP_SOC}% article
 \textsc{D.~Hobbs},  \textsc{G.~Kresse},  and  \textsc{J.~Hafner}\iffalse Fully
  unconstrained noncollinear magnetism within the projector augmented-wave
  method\fi,
 \jr{Phys. Rev. B} \textbf{62}(Nov), 11556--11570 (2000).


\bibitem{siesta_SOC_2004}% article
 \textsc{V.\,M. García-Suárez},  \textsc{C.\,M. Newman},  \textsc{C.\,J.
  Lambert},  \textsc{J.\,M. Pruneda},  and  \textsc{J.~Ferrer}\iffalse First
  principles simulations of the magnetic and structural properties of
  {Iron}\fi,
 \jr{The European Physical Journal B} \textbf{40}(4), 371--377 (2004).


\othercit
\bibitem{zerothi_sisl}% misc
 \textsc{N.\,R. Papior},
sisl: v0.8.5.


\end{thebibliography}
\end{document}